# A polymer-based technique to remove pollutants from soft contact lenses

Katherine Burgener[1], M. Saad Bhamla[1*]

[1]School of Chemical and Biomolecular Engineering, Georgia Institute of Technology, Atlanta, Georgia, United States of America

[*]saadb@chbe.gatech.edu (corresponding author)

**Abstract**

Purpose: To demonstrate an alternative to the rinse and rub (RR) method for cleaning pollutants from the exterior surface of soft contact lenses. This proposed technique is termed Polymer on Polymer Pollutant Removal (PoPPR), which utilizes the elastic properties of polydimethylsiloxane (PDMS) to physically remove contaminants from contact lens surfaces through non-adhesive unpeeling.

Methods: Three different ratios of setting agent to polymer PDMS (1:30, 1:40, and 1:50) were evaluated using the PoPPR method against the control method of RR with a commercial multi-purpose lens cleaning solution. Three simulated pollutants of different sizes: pollen (25-40 µm), microbeads (1-5 µm), and nanoparticles (5-10 nm), were used to test the effectiveness of both cleaning methods. The fraction of pollutants removed from each contact lens was recorded and evaluated for significance.

Results: PDMS 1:40 was found to be the optimal ratio for lens cleaning using the PoPPR method. For larger particles (>10 µm), no difference was observed between conventional RR and proposed PoPPR method ($p > 0.05$). However, the new PoPPR technique was significantly better at removing small $PM_{2.5}$ particles (<2.5 µm) compared to the RR method, specifically for microbeads ($p = 0.006$) and nanoparticles ($p < 0.001$).

Conclusion: This proof-of-concept work demonstrates that the PoPPR method of cleaning contact lenses is as effective as the conventional cleaning method for larger particles such as pollen. The PoPPR method is more effective at removing extremely fine particulate pollutants, including microplastics and nanoparticles. This method offers a potentially more efficient cleaning protocol that could enhance the safety, health, and comfort of contact lens users, especially those living in regions with significant air pollution.

**Introduction**

Solid pollutants are classified by size as particulate matter (PM). $PM_{2.5}$ particles are airborne pollutants with diameters less than 2.5 µm. Both natural and manufactured air pollutants fall within this size range; common examples of these micro and nano-pollutants are carbon-rich fluffy soot aggregate, cigarette smoke, pollen, mushroom spores, particulates with metallic elements, as well as diesel exhaust **(Fig.1)** [1,2]. $PM_{2.5}$ pollutants are relevant to ocular health because they decrease tear osmolarity [3] and increase dry eye, ocular irritation, and burning [4].

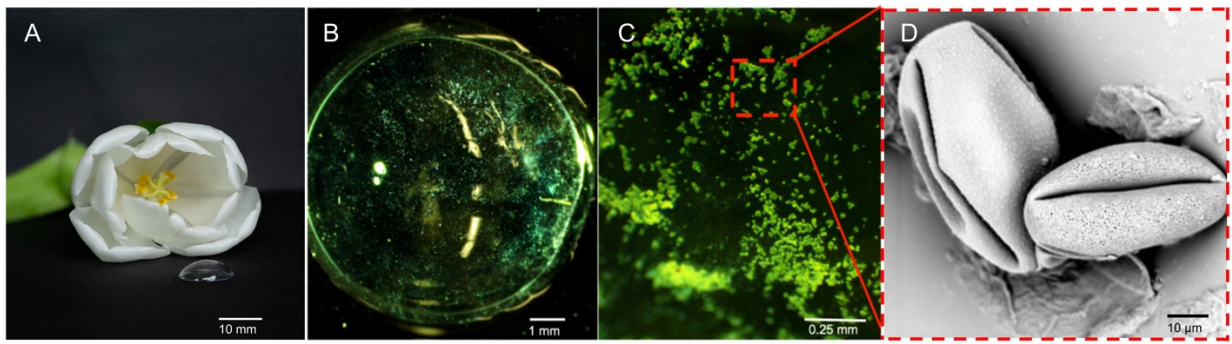

**Figure 1: A.** Pollen is both a common allergen and airborne pollutant. For the general population, pollen on a contact lens may cause irritation, swelling, and scratching of the lens and upper eye lid. Further, pollen that sticks to an allergic person's lenses can cause prolonged allergic reactions to occur. Contact lenses are exposed to airborne pollutants during open eye lens wear. **B.** 1x image of pollen on the surface of a contact lens with a stereo microscope. **C.** 40x image of pollen taken with a stereo microscope. **D.** Pollen from the same source (tulip) can exhibit different geometries, image taken with an SEM microscope.

Typical soft contact lens pore sizes are on the order of sub-micrometers, with diameters up to 0.2 µm [5]. Large foreign objects (lashes, mascara, debris) do get caught on contact lenses and the conventional cleaning method of rinsing with cleaning solution and rubbing with fingers (RR) appears to work in dislodging them from the hydrogel matrix [6]. Due to their rough nature, smaller bodies (minerals, aerosol pollution) have the potential to embed themselves in the lens, rendering rubbing inadequate and potentially harmful to the surface of the lens [7].

Recent improvements in the hydrogel polymer matrix of contact lenses have increased the lenses' antimicrobial properties but have not changed the way contact lenses are cleaned (i.e. rubbing the lens or using multi-purpose solutions (MPS)) [8,9,10]. Some MPS were developed as a 'no-rub' cleaning alternative that disinfected lenses by rinsing and extended soaking overnight. While this cleaning technique has been shown to decrease biofilm activity on lenses, not all MPS solutions are able to remove physical deposits from the lens [11]. Rubbing has been demonstrated as a necessary addition to cleaning regimens due to the ability of the shear forces applied by the fingers to remove pollutants [6]. Prolonged wear of lenses leads to physical fatigue of the lens integrity [12]; constant mechanical stress changes the morphology of the surface and the pore sizes of the lenses [13,14]. The repetitive motion of blinking can damage the lens, so the shear forces of fingers rubbing across the surface would be expected to also damage the surface of lenses.

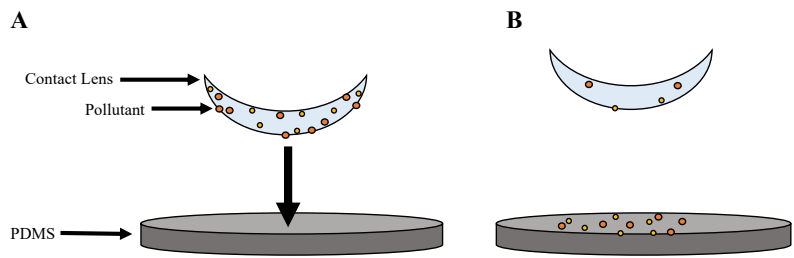

**Figure 2**: Polymer on polymer pollutant removal (PoPPR) is the proposed method of cleaning. **A.** The contact lens surface is gently pressed against PDMS using a finger. **B.** As the contact lens is removed, pollutants stick to the PDMS.

Polydimethylsiloxane (PDMS) has many uses due to its physical properties of straightforward fabrication and transparency, though it is used most frequently in microfluidic design and research [15]. PDMS was chosen as the candidate material in this experiment due to its extensive use and testing in research and industry. PDMS is widely and commercially available and its properties are well studied and understood. More specifically, when set, PDMS is a soft, inert, and elastic material with tunable Young's modulus [16], low autofluorescence [17], and excellent biocompatibility [19]. The proposed new method of contaminant removal takes advantage of these ideal characteristics. Polymer on polymer pollutant removal (PoPPR) involves pressing a contaminated contact lens onto a surface of PDMS. The PDMS envelops the pollutants on the lens while the hydrophobicity of PDMS prevents the lens from sticking to the PDMS surface. The result is a transfer of pollutants from lens to PDMS, as shown in **Fig. 2**. The aim of these proof-of-concept experiments was to determine how effective PoPPR is at cleaning physical pollutants from contact lens surfaces and to compare the results to control method, RR.

**Materials & Methods**
Contact Lenses
All contact lenses used for experiments were 1-Day Acuvue TruEye +6.0 lenses (Johnson & Johnson Vision, Jacksonville, FL). The lenses were rinsed with Biotrue multi-purpose solution (Bausch & Lomb, Rochester, NY) before running experiments.

PDMS
SYLGARD™ 184 Silicone Elastomer was used to make the PDMS (2646340, Dow). The commercial Sylgard kit contains both the PDMS polymer and the elastomer curing agent. The product suggests mixing a 1:10 ratio of setting agent to polymer. To increase the elasticity of the PDMS, the following ratios were used: 1:30, 1:40, 1:50. Both liquids were combined in 35 mm petri dishes. The average total weight of each experimental sample was approximately 18 g. To ensure an even distribution of setting agent in the polymer, samples were stirred vigorously for five minutes. After mixing the two liquid parts, samples were de-aired using a vacuum desiccator until all bubbles were removed from the mixture. Each sample was cured for at least 24 hours at 50°C in an incubator at 10% humidity.

Pollutants
*Populus Tremuloids* pollen averages 25-40 µm in diameter and was obtained for pollen removal trials (P7770-500MG, Sigma Aldrich). Fluorescent microbeads with diameters between 1-5 µm were used for the microbead removal experiments (300-45-225, Cospheric). For nanoparticle removal trials, copper indium disulfide/zinc sulfide quantum dots between 5-10 nm in diameter were used (29-8520, Strem Chemicals). All pollutants were suspended in water before application to contact lenses.

Cleaning Experiments

The following experimental procedure was used.

**Part a. Controlled Fouling of Lenses**
First, clean lenses were polluted with a solution of water and pollutant. The concentration of pollen in water was 0.001 g/mL; the microbead solution had a concentration of $1.225 * 10^{-5}$ g/mL. Pollen and Microbead: Both pollen and microbeads were aerosolized onto the lens with a spray bottle, which released approximately 1 mL of solution every five sprays. Nanoparticle: 0.5 mL of $5.4*10^{-5}$ g/mL nanoparticle solution was pipetted onto the lens in order to avoid harmful aerosolization. The concentration of particles deposited on lenses using aerosolization (**Fig. 1. B**) or pipetting (**Supplemental Information (SI) Fig. 2**) were distributed throughout the entire lens area as evidenced through visual inspection using a microscope.

**Part b. Cleaning Protocol**
i). RR protocol: The lenses were cleaned by rinsing with standard multi-purpose cleaning solution and rubbing the polluted surface between the thumb and index finger (see **SI Fig. 1**). Since RR is a qualitative procedure, to maintain consistency between trials, the lens was gently rubbed for five seconds, followed by a rinse for five seconds. The rinsing solution was put in sterile petri dishes. Great care was taken to ensure that all cleaning solution was collected.
ii). PoPPR protocol: The polluted surfaces of lenses were pressed into a sample of clean PDMS. Lenses were only pressed once onto the PDMS surface and peeled away after five seconds of contact.

**Part c. Analysis**
After the cleaning step, the contact lenses were allowed to dehydrate (24h) and were then sandwiched between two clean glass slides to enable viewing under a microscope. The rinsing solution from RR protocol was allowed to evaporate so that the pollutants were left on the bottom of the petri dishes. Images were taken of the used lens, and either the rinsing solution petri dish or PDMS with a fluorescence microscope (**SI Fig. 2**). To image the entire surface at 10x magnification, approximately 13,15, and 30 images were required of the contact lens, petri dish, and PDMS respectively.

Using the 'analyze particles' program on Fiji [19], pollutants were counted, and the fraction of pollutant removal was calculated. The fraction of pollutant removed was calculated as follows:

$$\text{Fraction Removed} = \frac{n_{removed}}{(n_{removed} + n_{remaining})}$$

where $n_{removed}$ refers to the number of pollutant particles either on the PDMS or in the petri dish and $n_{remaining}$ is the number of pollutant particles on the used lens. It should be noted that the initial number of particles deposited using the controlled fouling step is not measurable as it involves drying the lens to manually count, which renders the cleaning protocol difficult. Thus, all cleaning experiments were performed carefully so that $n_{removed} + n_{remaining}$ could be used as a proxy for the total initial particles.

The cleaning efficiencies of PoPPR and the control (RR) were compared for three different pollutant sizes (25-40 μm, 1-5 μm, 5-10 nm). For all pollutant sizes, at least four trials were conducted for each test to ensure repeatability and reproducibility. Because pollen is not fully fluorescent and its shape is not uniform, each pollen grain on the lens, in solution, or on PDMS, was hand-counted. For this reason, only four trials were conducted. Experimental microbeads are fully fluorescent and were able to be counted with Fiji software, ten trials were deemed sufficient for this proof-of-concept paper. Nanoparticles are also fluorescent but due to the risks of aerosolization, only five trials were performed for each cleaning method.

To examine the role of PDMS stiffness in cleaning efficiency for PoPPR, three different ratios of PDMS were also tested (1:30, 1:40, and 1:50).

Statistical analysis
Statistical analyses were performed using Statistical Analysis Software (SAS) Version 9.4 (SAS Institute, Cary, NC). The Shapiro-Wilke test was used to assess normality of the data. Wilcoxon (Rank Sums) tests were performed on each pollutant size to assess differences in effectiveness between the cleaning methods. Outliers were identified using the Grubbs test. Standard deviations ($\sigma$), means ($\mu$), mid-ranges (MR), and number of trials ($n$) for each cleaning method were recorded. Finally, to assess statistical significance between cleaning methods, two-tailed two-sample t-tests were conducted at a 95% level of significance. The means and standard deviations are presented as $\mu \pm \sigma$ in the results section below.

**Results**
Pollen Removal
The fraction of pollen removed using the control method (RR) averaged $0.57 \pm 0.2$ ($n = 4$). The mean fractions of pollen removed from PDMS 1:30, 1:40 and 1:50 using the PoPPR method were $0.48 \pm 0.01$ ($n = 4$), $0.65 \pm 0.07$ ($n = 4$), and $0.62 \pm 0.14$ ($n = 5$) **(Fig. 3 A)**.

For the removal of pollen from contact lenses, there were no statistical differences in the cleaning methods ($p = 0.249$). The only statistical difference in pollen removal methods was between PDMS 1:30 and 1:40 PoPPR ($p = 0.0024$).

Microbead Removal
The mean fraction of microbeads removed from lenses using the RR method was $0.50 \pm 0.16$ ($n = 10$). For the PoPPR trials of PDMS 1:30, 1:40 and 1:50, the mean fractions of pollutant removed were $0.43 \pm 0.06$ ($n = 5$), $0.67 \pm 0.09$ ($n = 10$), and $0.67 \pm 0.11$ ($n = 10$) respectively (**Fig. 3 B**).

Statistical differences were observed between RR and PDMS 1:40 ($p = .006$) as well as RR and PDMS 1:50 ($p = .011$). PDMS 1:30 results were also different from PDMS 1:40 ($p < 0.001$) and 1:50 PDMS ($p < 0.001$). In the microbead data for PDMS 1:40, there was one significant outlier, but removing the outlier did not change statistical results.

Nanoparticle Removal
The RR method resulted in a mean fraction of $0.37 \pm 0.18$ ($n = 5$) for nanoparticle removal. The mean fractions of nanoparticles removed from lenses using the PoPPR methods of PDMS 1:30, 1:40, and 1:50 were $0.40 \pm 0.15$ ($n = 5$), $0.85 \pm 0.06$ ($n = 5$), and $0.70 \pm 0.10$ ($n = 5$) (**Fig. 3 C**).

The following cleaning methods were statistically different: RR and PDMS 1:40 ($p < 0.001$), RR and PDMS 1:50 ($p = 0.007$), PDMS 1:30 and PDMS 1:40 ($p < 0.001$), PDMS 1:30 and PDMS 1:50 ($p = 0.0055$), PDMS 1:40 and PDMS 1:50 ($p = 0.021$).

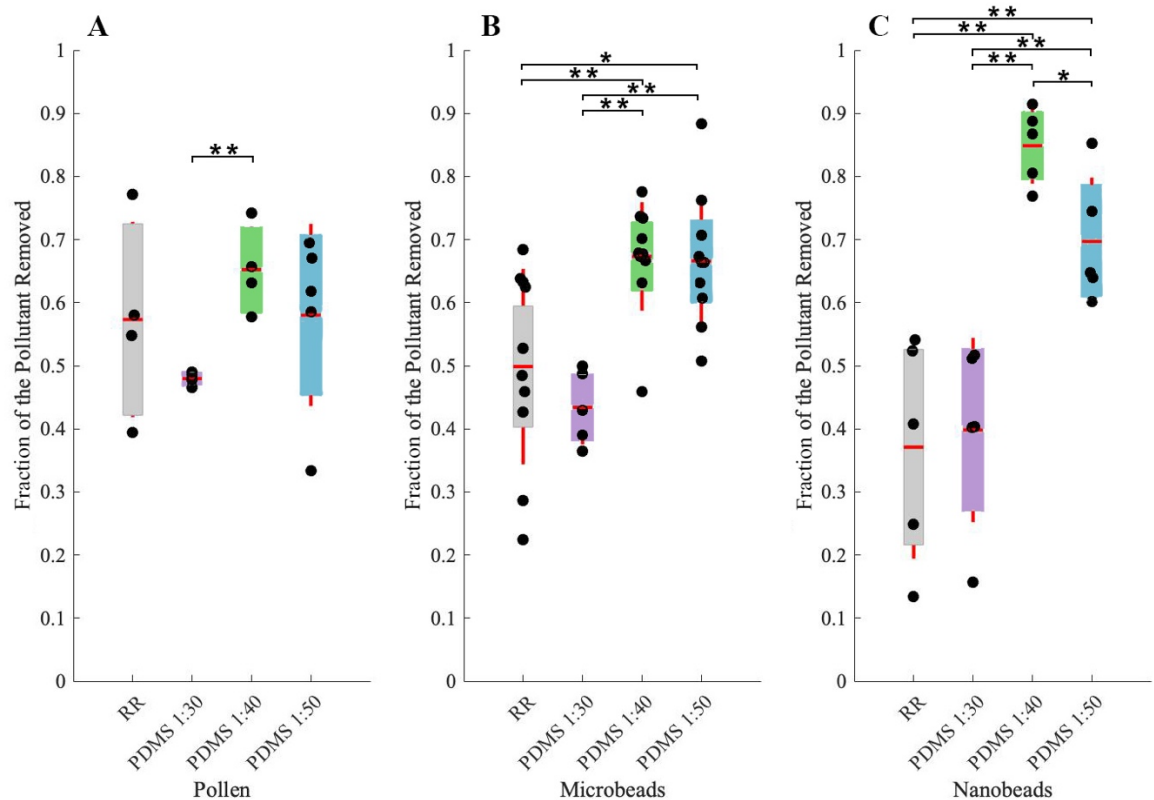

**Figure 3**: Data collected includes the fraction of pollutant removed using four different cleaning protocols. The horizontal red lines indicate the mean of each data set, the colored bars represent one standard deviation, and the vertical red lines show the 95% confidence level. **A.** Results from pollen removal trials (n>4). **B.** Microbead removal data (n>5). **C.** Results from nanoparticle removal trials (n=5). * $p < .03$, ** $p < .01$.

## Discussion

The PoPPR method was comparable to the RR method for pollen pollutants. **Fig. 3** shows that the means ($\mu$) and standard deviations ($\sigma$) of the RR method. The large mid-range (MR) in RR data (0.2) compared to a smaller MR of PDMS 1:40 data (0.08) indicates that the RR method is not as precise as the PoPPR technique.

The large size of the experimental pollen can explain some of these results. *Populus Tremuloids* pollen has an average diameter of 25-40 µm which is larger than the average contact lens pore size (< 1.0 µm). While it is possible that some pollen was embedded in the contact lens pores, most of the pollutant likely rested on the surface of the lens (**Fig. 1**). Results suggest that for large particles (>25 µm), PoPPR is as effective as the RR method.

Conversely, the RR and PoPPR methods differed statistically. **Fig. 3** provides a visual representation of results. Both PDMS 1:40 and 1:50 were statistically better at removing microbeads from contact lens surfaces than the RR method and PDMS 1:30. Similar to the pollen results, the RR data exhibited large values of $\sigma$ and MR, indicating large variability in this technique which is expected due to the qualitative nature of this protocol.

Nanoparticle cleaning data show the largest improvement in cleaning method performance. The PDMS 1:40 PoPPR was significantly better at cleaning nanoparticles than RR. Similar to pollen and microbeads, RR and PDMS 1:30 methods remove an approximately equal fraction of nanoparticles.

In summary, the PDMS 1:40 PoPPR method removed more pollutants than the RR method (**Fig. 4**). The control method (RR) and the proposed PoPPR technique were comparable when considering large pollutants (> 25 μm). However, the PDMS 1:40 was significantly more successful at small contaminant removal ($PM_{2.5}$) such as microplastics and nanoparticles. During RR removal, the frictional or shearing forces that act on the lens surface are assumed to dislodge particles. These forces may be sufficient for detaching larger particles (pollen), however, for nanoparticles, these forces may not be sufficient. This is attributed further to the rough contact lens surface at the nanometer length scale, where the nanoparticles might be tightly adsorbed on the lens surface. In contrast, the PoPPR technique leverages a normal or extensional peeling force applied to physically detach particles, which works across all size ranges and is surprisingly effective for smaller particles.

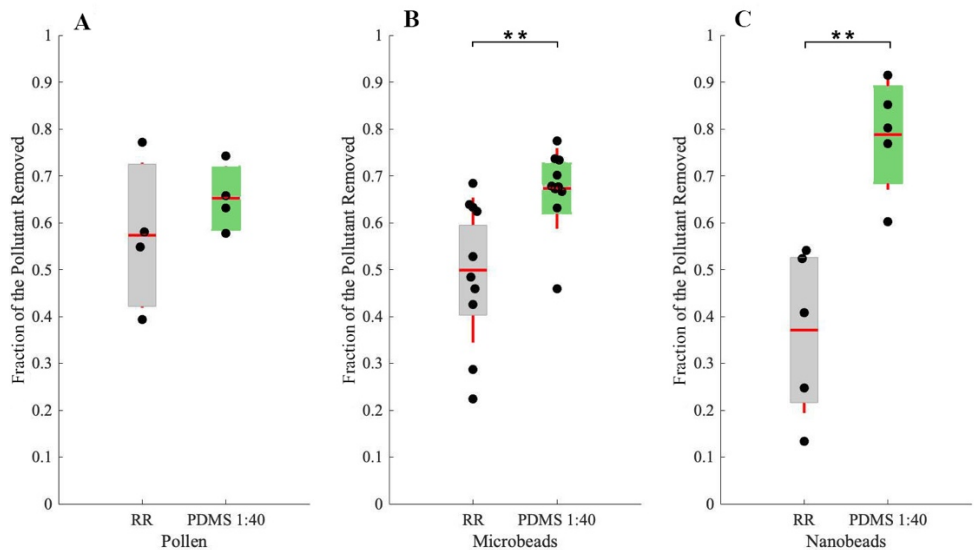

**Figure 4:** Results comparing RR to PoPPR cleaning methods for three different sized contaminants. When compared with a two-tailed t-test, the PoPPR method using PDMS 1:40 is significantly more effective at removing microbead and nanoparticle pollutants. **A.** Results from pollen removal trials. **B.** Microbead removal data. **C.** Results from nanoparticle removal trials. ** $p < .01$.

The results also suggest that an optimal PDMS stiffness exists that enhances particulate removal from lenses using the PoPPR technique. While the PDMS 1:30 results indicate that there is no statistical significance to RR, 1:30 results are statistically inferior to their PDMS 1:40 counterparts for all pollutants. These results are expected as higher ratio PDMS (≥1:40) are softer (lower Young's modulus) than 1:30, enabling the matrix of PDMS to surround small particles and detach them during peeling [20]. However, the underlying particle removal mechanism based on the PDMS stiffness remains an open question and will be the focus of future work.

A major reason why contact lens cleaning techniques have not changed is that the current method is inexpensive, rapid, and convenient. While rinsing and rubbing is effective for large contaminants, it may negatively impact lens surface integrity (tears, scratches) [21, 22]. Additionally, results of the RR method can be highly variable depending on the user [6]. In contrast, the proposed PoPPR technique offers a more efficient, effective, and repeatable pollutant removal technique that utilizes a widely available, inert, and biocompatible polymer (**SI Fig. 3**). There are, however, a few shortcomings of the PoPPR method and limitations of this current study. The two shortcomings of the PoPPR method include the introduction of an extra step in cleaning routines and the potentially higher cost, both of which may be unattractive to consumers. The limitations of this current work are first, all the tests done here are for one type of commercial silicone hydrogel contact lens, as a proof-of-concept of this new technique. Thus, to generalize these results, future work will focus on extensive comparison between the different types of contact lens materials and a huge variety of commercial cleaning solutions, which is beyond the focus of this study. Second, additional testing needs to be done to test the efficacy of the PoPPR technique in removing tear film lipid and protein deposits that are often the more common fouling concerns in contact lens wearers. Despite these shortcomings and limitations, the proof-of-concept cleaning technique presented here has the potential to improve the comfort and long-term use of contact lenses, especially for users in regions with heavy air pollution.

**Conclusion**
This proof of concept project successfully demonstrated that the PoPPR method for cleaning contact lenses contaminated with different sized contaminants was comparable to or more effective than the traditional method of rinsing and rubbing with lens cleaners. PDMS that has a setting agent to polymer ratio of 1:40 and 1:50 was determined to be the most effective at removing pollutants of all experimental sizes.

**Declarations of Interest:** None

**Funding Source:** None

# A polymer-based technique to remove pollutants from soft contact lenses

Katherine Burgener[1], M. Saad Bhamla[1*]

[1]School of Chemical and Biomolecular Engineering, Georgia Institute of Technology, Atlanta, Georgia, United States of America

[*]saadb@chbe.gatech.edu (corresponding author)

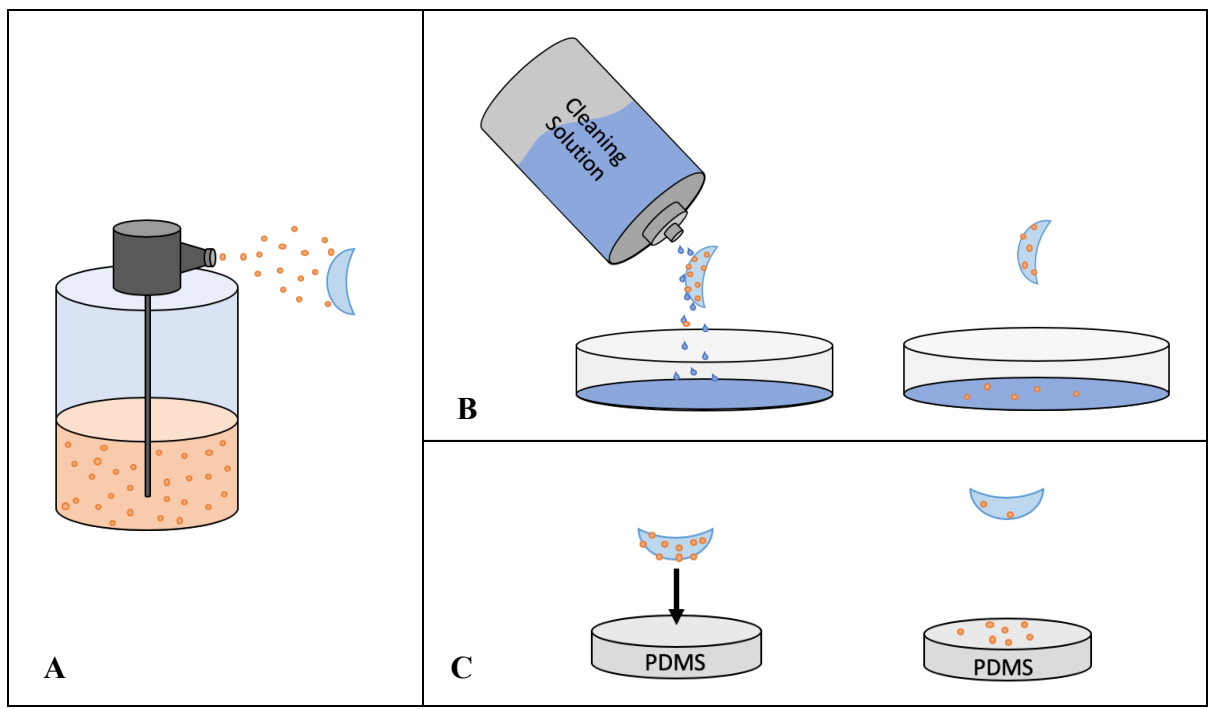

**Supplemental Figure 1**: Experimental method for cleaning contaminated contact lenses. **A.** For both cleaning methods pollutants suspended in water were sprayed onto a contact lens. **B.** Rinsing a lens. After rubbing the contact lens between the thumb and the forefinger (not pictured), lenses are rinsed with cleaning solution. The pollutants on the contact lens and the pollutants in the solution are counted. **C.** PoPPR cleaning method. A polluted lens is pressed onto a sample of PDMS with normal force, then lifted off the surface. The pollutants on the contact lens and on the PDMS are counted.

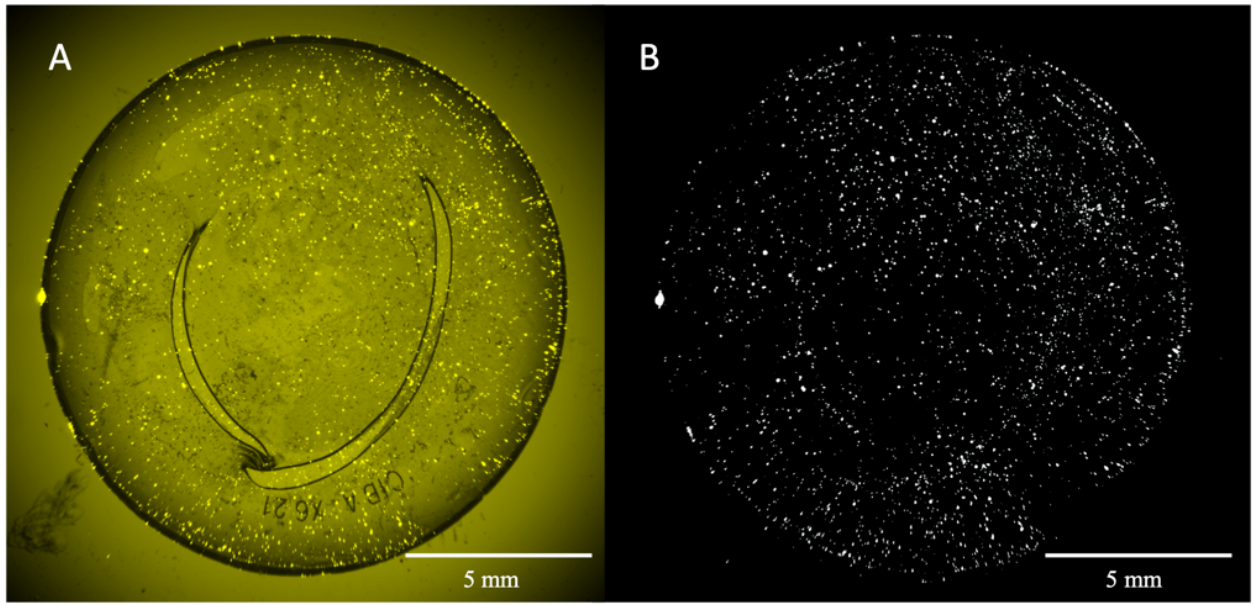

**Supplemental Figure 2**: Fluorescent microbeads are easily discernable under a microscope at 5x magnification. For accurate data analysis, the contact lens was magnified 25x and pictures were taken successively to capture the entire lens. **A.** Fluorescent beads on the surface of contact lens under a combination of fluorescent and natural lighting. **B.** Same lens under complete fluorescence.

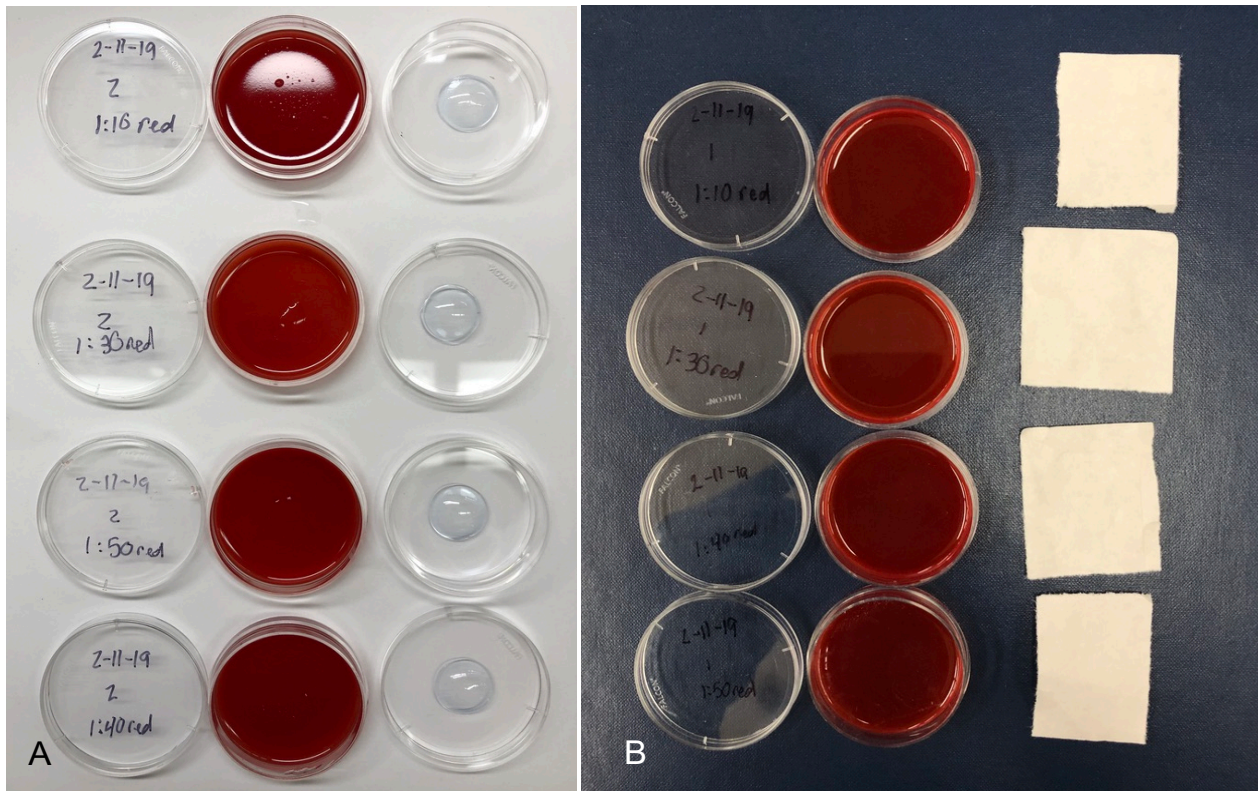

**Supplemental Figure 3**: Experimental analysis of PDMS leaving residue. PDMS polymer was dyed before adding the setting agent then cured as normal. Four different ratios of setting agent to polymer were tested: 1:10, 1:30, 1:40, and 1:50. **A.** Residue testing on contact lenses. Clean contact lenses were pressed onto the PDMS samples then removed. No PDMS was visible on the lenses microscopically. **B.** Residue testing on paper. Unlike the contact lenses, paper is completely dry and has different surface properties, mainly high absorbance. Clean white paper samples were pressed onto the PDMS samples and removed. No PDMS was visible on the paper.